\documentstyle[epsfig,12pt]{article}
\input axodraw.sty
\newcounter{minutes}
\def\lsim{\:\raisebox{-0.5ex}{$\stackrel{\textstyle<}{\sim}$}\:}
\def\gsim{\:\raisebox{-0.5ex}{$\stackrel{\textstyle>}{\sim}$}\:}
\title{\vskip-3cm{\baselineskip16pt
\centerline{\normalsize DESY 97--242\hfill ISSN 0418-9833}
\centerline{\normalsize BI-TP 97/56 \hfill}
\centerline{\normalsize {\tt hep-ph/9712309}\hfill}
}
\vskip1.5cm
\bf Production of Hard Photons and Jets \\
in Deep Inelastic Lepton Proton Scattering \\
at Order $O(\alpha_s)$}
\date{}
\author{ { G.\ Kramer$^a$, 
           D.\ Michelsen$^a{}^{*}$, 
           H.\ Spiesberger$^b{}$} \\[2ex]
  {\normalsize $^a$II. Institut f\"ur Theoretische
    Physik${}^{\dagger}$, Universit\"at Hamburg}\\
  {\normalsize D - 22761 Hamburg, Germany}\\[1ex]
  {\normalsize $^b$Fakult\"at f\"ur Physik${}^{\dagger}$,
    Universit\"at Bielefeld}\\
  {\normalsize D - 33615 Bielefeld, Germany}\\[1ex]}
\begin{document}
\newcommand{\newc}{\newcommand}
\newc{\gjo}{$\gamma+(1+1)$-jet}
\newc{\gjt}{$\gamma+(2+1)$-jet}
\maketitle
%
%
\vspace{2cm}
\begin{abstract}
  We calculate the $O(\alpha_s)$ corrections to the production of a
  hard and isolated photon accompanied by one or two jets in deep
  inelastic lepton nucleon scattering at HERA. Numerical results are
  presented and the potential of this process for studies of parton
  distribution functions is discussed. 
\end{abstract}
\vspace*{\fill}

\footnoterule

{\footnotesize
\noindent
${}^*$ Now at IBM Deutschland, Informationssysteme GmbH. \\
${}^{\dagger}$ Supported by Bundesministerium f\"ur Bildung, 
  Wissenschaft, Forschung und Technologie, Bonn, Germany, Contracts 
  05 7BI92P (9) and 05 7HH92P (0).}

\newpage

\section{Introduction}

The production of hard photons in hadronic processes is an important
testing ground for QCD. Since the photon does not take part in the
strong interaction, it is a 'direct' probe of the hard scattering
process. Direct photon production in $\gamma p$ \cite{gammap} and in
$p\bar{p}$ collisions \cite{ppbar} provides a means to determine the
strong coupling constant $\alpha_s$ and has been used to extract
information on the parton distributions, in particular the gluon
density in the proton \cite{gluon}. In $e^+e^-$ annihilation
\cite{lep}, measurements of photon radiation in hadronic $Z$ decays at
LEP1 have provided important independent information on the
electroweak couplings of {\it up} and {\it down} quarks to the $Z$
boson \cite{zcoupl,zcouple}.  Moreover, final states containing a
photon are an important background for many searches for new physics
and a good knowledge of the standard model predictions for direct
photon production is therefore required.

At HERA, radiative deep inelastic scattering, $ep \rightarrow e\gamma
X$, with photons collinear to the incoming electron has been used to
obtain a measurement of the structure function $F_2$ at low values of
the momentum transfer $Q^2$ \cite{radscatt}. Also the first
observation of hard non-collinear photons at $Q^2 = 0$, {\it i.e.}\ in
photoproduction has been reported recently \cite{zeusph}. With
increasing luminosity this measurement is expected to contribute
information on the parton content of the photon and the proton. By
contrast, direct photon production at large $Q^2$ would be sensitive
to the parton distributions in the proton only.  The information
obtained this way would be complementary to the $F_2$ measurement from
inclusive deep inelastic scattering, since {\it up} and {\it down}
quarks contribute with different weights. Typical cross sections for
the production of hard photons in deep inelastic scattering with $Q^2
> 10$ GeV${}^2$ are of the order of 10 pb. With a luminosity of 50
pb${}^{-1}$ one thus expects statistical uncertainties of the order of
5\,\% and a measurement of differential cross sections seems feasible.

Whereas next-to-leading order calculations for direct photon
production are available for photoproduction \cite{gammap,nlogammap},
$p\bar{p}$ collisions \cite{nloppbar}, as well as for $e^+e^-$
annihilation \cite{theo-lep1,theo-lep2}, a corresponding calculation
for deep inelastic $ep$ scattering was still missing.  In this work we
study the $O(\alpha_s)$ corrections to the process $ep \rightarrow
e\gamma X$ at large $Q^2$. Since hard photon production is a process
of relative order $\alpha_e = 1/137$ with respect to the total deep
inelastic scattering cross section, we expect sizable event rates only
at moderately large $Q^2$ and restrict ourselves therefore to pure
photon exchange, {\it i.e.}\ $Z$-exchange contributions are neglected.
The calculations will be organized in such a way that the hadronic
final state can be separated into \gjo\ and \gjt\ topologies (the
remnant being counted as ``$+1$" jet, as usual). Our approach is thus
analogous to that in calculations of $(2+1)$- and $(3+1)$-jet cross 
sections in deep inelastic scattering where a gluon is replaced by a 
photon \cite{epjets}.  \gjt\ events originate through the emission or 
absorption of a gluon. Therefore the ratio of \gjt\ and \gjo\ events 
is sensitive to the value of the strong coupling constant $\alpha_s$ 
and to the gluon distribution. 

In addition to perturbative direct production, photons are also
produced through the `fragmentation' of a hadronic jet into a single
photon carrying a large fraction of the jet energy \cite{fragm}. This
long-distance process is described in terms of the quark-to-photon and
gluon-to-photon fragmentation functions.  The necessity for taking
into account non-perturbative contributions is signaled by the
presence of singularities showing up in a perturbative calculation.
These singularities are related to collinear photon-quark
configurations. The factorization theorem of QCD guarantees that all
singularities can be absorbed into well-defined universal
parton-to-photon fragmentation functions, the remainder being
calculable in perturbation theory.

In practice, a measurement of direct photon production is feasible
only when isolation conditions are imposed on the observed photon in
order to reduce various hadronic backgrounds, in particular from
two-photon decays of $\pi^0$. The contribution from non-perturbative
parton-to-photon fragmentation, being related to collinear photon
emission from partons, can be reduced by isolation requirements, but
is not completely removed. Again, in a perturbative calculation, this
is related to the presence of singularities. In fact, if one tries to
model the experimental isolation conditions by imposing cutoffs on
parton-level jets, one can not exclude contributions due to soft
quarks having emitted a hard collinear photon; the soft quark may
appear only as part of a parton-level jet, but not as a separate,
observable jet which can enter the isolation conditions\footnote{The
  problem is most easily visible in $e^+e^- \rightarrow \gamma +
  1$-jet, where already at leading order photon-jet isolation does not
  remove the photon-quark collinear singularity \cite{theo-lep2}. A
  next-to-leading order calculation \cite{gehrmann} shows features
  typical for a next-to-next-to-leading order calculation.
  Measurements of the quark-to-photon fragmentation function in
  $e^+e^- \rightarrow \gamma + 1$-jet had been proposed in Refs.\ 
  \cite{morgan,bourhis} and were described in Refs.\ 
  \cite{frag-exp}.}.  The implementation of photon isolation is
particularly non-trivial in a calculation including $O(\alpha_s)$
contributions since the isolation conditions affect the available
phase space for gluon emission \cite{berger}. As a consequence, the
parton-to-photon fragmentation functions may have to be modified for
isolated photon production and higher-order corrections may turn out
to be large and to require their resummation.

In the present work we adopt a simpler approach where the
fragmentation contributions are ignored completely. The photon-quark
collinear singularities then have to be removed by explicit
parton-level cutoffs. The dependence of the final results on these
cutoffs (discussed in section 4 below) will indicate to what extent
the quark-to-photon fragmentation function would contribute in a more
systematic treatment.


\section{The Leading-Order Process}

In leading order (LO), the production of photons in deep inelastic
electron (positron) proton scattering is described by the quark
(antiquark) subprocess
\begin{equation}
e(p_1) + q(p_3) \rightarrow e(p_2) + q(p_4) + \gamma(p_5)
\label{LOprocess}
\end{equation}
where we have given the definition of the particle momenta in
parentheses. The momentum of the incoming quark is a fraction of the
proton momentum $p_P$: $p_3 = \xi p_P$. The proton remnant $r$ carries
the momentum
\begin{equation}
p_r = (1-\xi) p_P
\end{equation}
and hadronizes into the remnant jet so that the process
(\ref{LOprocess}) gives rise to \gjo\ final states.  The momentum of the
hadronic final state, {\it i.e.}\ the $(1+1)$-jet system, is $p_P + p_1 -
p_2 - p_5$ and its invariant mass $W$ is given by
\begin{equation}
W^2 = (p_P + p_1 - p_2 - p_5)^2.
\end{equation}
We will use the well-known kinematic variables for deep inelastic
scattering
\begin{equation}
Q^2 = - (p_1 - p_2)^2, ~~~~~
x = \frac{Q^2}{2p_P(p_1 - p_2)}, ~~~~~
y = \frac{Q^2}{xs},~~~~~
s = (p_1 + p_P)^2, 
\end{equation}
determined by the momentum of the scattered lepton. Because of the
presence of the photon in the final state, large $Q^2$ does not
guarantee large $W$ and we will have to require explicitly $W > W_{\rm
  min}$ in order to stay in the deep inelastic regime where a
perturbative treatment can be expected to work. Apart from this, we
will also apply cuts on the variables $x$, $y$ and $Q^2$ since we ask
for an observable scattered electron. These latter cuts remove direct
photon production in photoproduction.

Both leptons and quarks emit photons. The subset of Feynman diagrams
where the photon is emitted from the lepton (``leptonic radiation") is
gauge invariant and can be treated separately. Similarly, the Feynman
diagrams with a photon emitted from the quark line is called
``quarkonic radiation". There is also a contribution from the
interference of these two parts. For tests of QCD the interest is in
those contributions where the photon is emitted from quarks and
leptonic radiation is viewed as a background.

Radiative deep inelastic scattering appears as a contribution to QED
radiative corrections (see for example Ref.\ \cite{radcorr} and
references therein). In this case the emitted photon remains undetected
and singularities due to soft and collinear photons have to be canceled
by taking into account virtual $O(\alpha)$ corrections to non-radiative
scattering $eq \rightarrow eq$. Here we are interested in events with an
observable photon, {\it i.e.}\ we restrict ourselves to the case where
the energy of the photon $E_{\gamma} = E_5$ is sufficiently large,
\begin{equation}
E_{\gamma} > E_{\gamma,{\rm min}}.
\end{equation}
Also, the photon should be spatially separated from all other
particles: 
\begin{equation}
\theta_{\gamma,i} > \theta_{\rm sep},
\end{equation}
where $\theta_{\gamma,i}$ is the angle between the momenta of the
photon and particle $i$ ($=1, 2, 3, 4$ for the leading-order process
(\ref{LOprocess}) and similarly for the next-to-leading order processes
specified in section 3 below). In particular, the photon is
not allowed to be emitted close to the beams:
\begin{equation}
\theta_{\rm min} < \theta_{\gamma} < \theta_{\rm max}.
\end{equation}
These cuts remove all photonic infrared and collinear singularities.
Instead of using the angle $\theta_{\gamma,i}$, photon separation from
final state particles can also be imposed by cuts on the invariant
masses
\begin{equation}
s_{ij} = (p_i + p_j)^2
\end{equation}
or, normalized to the invariant mass of the hadronic final state, 
\begin{equation}
y_{ij} = \frac{s_{ij}}{W^2}.
\end{equation}
The condition 
\begin{equation}
y_{5i} > y_0^{\gamma}
\label{y0g}
\end{equation}
($i=2, 4, r$) is more comfortable for the analytic calculation, but
less suited to experimental requirements. Since we will perform the
phase space integration with the help of Monte Carlo techniques, we
are not restricted to one specific choice of isolation criteria, but
we can apply a combination of the above cuts as will be described
below. 

At lowest order, each parton is identified with a jet and
photon-parton isolation corresponds to the isolation of the photon
from an observable jet. With isolation cuts, parton-to-photon
fragmentation does not contribute at this order.


\section{$O(\alpha_s)$ Corrections}

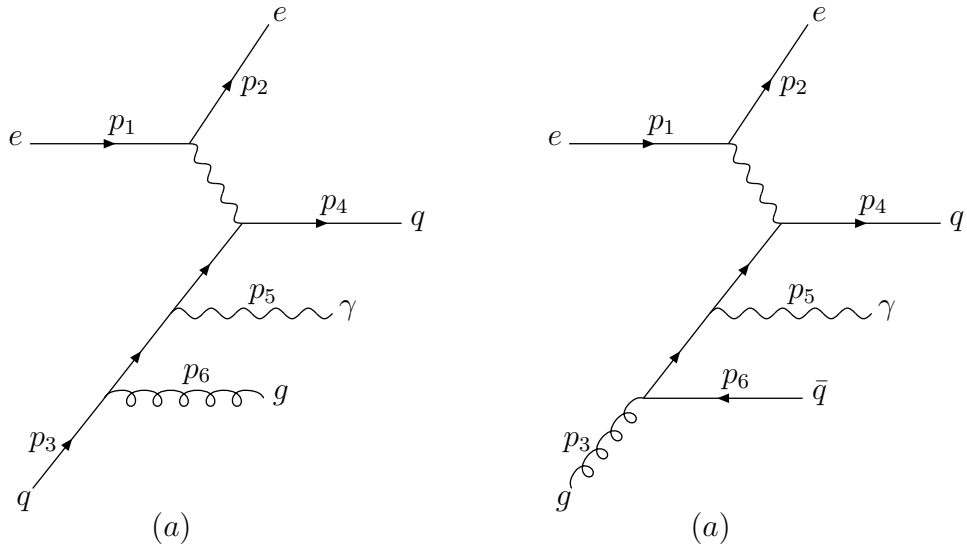
\begin{figure}[htbp]
%
\begin{picture}(200,200)(-30,0)
\ArrowLine(40,150)(100,150)
\ArrowLine(100,150)(130,195)
\Photon(100,150)(120,120){2}{4}
\ArrowLine(41,20)(68,54)
\ArrowLine(68,54)(93,86)
\ArrowLine(93,86)(120,120)
\ArrowLine(120,120)(180,120)
\Photon(93,86)(154,86){2}{5}
\Gluon(68,54)(128,54){3}{5}
\put(32,150){$e$}
\put(132,197){$e$}
\put(35,14){$q$}
\put(184,120){$q$}
\put(157,86){$\gamma$}
\put(132,54){$g$}
\put(70,156){$p_1$}
\put(120,171){$p_2$}
\put(150,126){$p_4$}
\put(123,92){$p_5$}
\put(98,62){$p_6$}
\put(40,36){$p_3$}
\put(86,2){$(a)$}
\end{picture}
%
\begin{picture}(200,200)(-30,0)
\ArrowLine(40,150)(100,150)
\ArrowLine(100,150)(130,195)
\Photon(100,150)(120,120){2}{4}
\Gluon(41,20)(68,54){3}{4}
\ArrowLine(128,54)(68,54)
\ArrowLine(68,54)(93,86)
\ArrowLine(93,86)(120,120)
\ArrowLine(120,120)(180,120)
\Photon(93,86)(154,86){2}{5}
\put(32,150){$e$}
\put(132,197){$e$}
\put(35,14){$g$}
\put(184,120){$q$}
\put(157,86){$\gamma$}
\put(132,54){$\bar{q}$}
\put(70,156){$p_1$}
\put(120,171){$p_2$}
\put(150,126){$p_4$}
\put(123,92){$p_5$}
\put(98,60){$p_6$}
\put(38,36){$p_3$}
\put(86,2){$(a)$}
\end{picture}
\caption{\label{Feynman} \it Examples of Feynman diagrams for $eq
  \rightarrow eqg\gamma$ (a) and $eg \rightarrow eq\bar{q}\gamma$ (b)
  with the definition of momenta.}
\end{figure}

At next-to-leading order (NLO), processes with an additional gluon,
either emitted into the final state or as incoming parton, have to be
taken into account:
\begin{equation}
e(p_1) + q(p_3) \rightarrow e(p_2) + q(p_4) + \gamma(p_5) + g(p_6),
\label{hoprocess1}
\end{equation}
\begin{equation}
e(p_1) + g(p_3) \rightarrow e(p_2) + q(p_4) + \gamma(p_5) + \bar{q}(p_6),
\label{hoprocess2}
\end{equation}
where the definition of momenta is again shown in parentheses (see
Fig.\ \ref{Feynman}).  In addition, virtual corrections (one-loop
diagrams at $O(\alpha_s)$) to the process (\ref{LOprocess}) have to be
included.

The amplitude for purely leptonic radiation at order $O(\alpha_s)$
factorizes into a leptonic tensor for $e \rightarrow e \gamma \gamma^*$
and a hadronic tensor including next-to-leading order QCD corrections.
Both parts are well-known and their combined contribution to deep
inelastic scattering is included for example in the Monte Carlo program
DJANGO6 \cite{django6}. For the $O(\alpha_s)$ corrections to quarkonic
radiation and in particular the lepton-quark interference, a
representation in terms of a leptonic and a hadronic tensor is not
suitable. The corresponding complete matrix elements including the
leptonic and hadronic vertex have been obtained with the help of {\tt
  form} \cite{form} and are given in \cite{michelsen}.

Whereas the LO process leads to the appearance of events with a photon
and one current jet, \gjo s, in higher orders additional jets can be
produced: the processes (\ref{hoprocess1}, \ref{hoprocess2})
contribute both to the \gjo cross section, as well as to the cross
section for \gjt s, depending on whether the quark-gluon or
quark-antiquark pair in the final state appears as one single jet or
as two separated jets.  The two cases can be identified by comparing
the scaled invariant masses of parton pairs with a jet resolution
parameter $y^J$: two partons $(i,j)$ with $i, j = 4, 6, r$ are
supposed to lead to 2 jets if
\begin{equation}
y_{ij} > y^J.
\end{equation}
Also the remnant $r$ is treated as a parton and a quark, antiquark, 
or gluon in the final state is recombined with the remnant into one 
jet if $y_{ir} = \frac{1-\xi}{\xi} y_{i3}$ is smaller than $y_0^J$. 
Similarly, photon isolation can be imposed with the help of cuts on 
the scaled invariant masses.

In the phase space region where two jets can not be separated, the
matrix elements become singular. These singularities appear when one
of the partons becomes soft or when two partons become collinear to
each other.  The singularities can be assigned either to the initial
state or to the final state (ISR: initial-state radiation, FSR:
final-state radiation).  The FSR singularities cancel against
singularities from virtual corrections to the lower-order process. For
the ISR singularities, this cancellation is incomplete and the
remaining singular contributions have to be factorized and absorbed
into renormalized parton distribution functions \cite{renorm}.

To accomplish this procedure, the singularities have to be isolated in
an analytic calculation, {\it e.g.}\ with the help of dimensional
regularization. The application of dimensional regularization is,
however, not feasible for the complete cross section of the
higher-order processes. Therefore we use the so-called phase-space
slicing method \cite{slicing} to separate those regions in the
4-particle phase space which give rise to singular contributions. A
separation cut $y_0^J$ is applied to the scaled invariant masses
$y_{ij}$ and chosen small enough, such that the calculation can be
simplified by neglecting terms of the order $O(y_0^J)$.  Contributions
from phase space regions where one of the $y_{ij}$ is smaller than
$y_0^J$ are singular and have to be combined with the one-loop
corrections to obtain a finite result. The sum of these two
contributions defines the cross section for events where two partons
are recombined into a parton-level jet (parton-level $(1+1)$-jet
events). The contributions where all $y_{ij}$ are bigger than $y_0^J$
are related to final states with three separate partons (parton-level
$(2+1)$-jet events). The latter are free of singularities and can be
calculated with the help of Monte Carlo techniques.

As known from similar calculations ({\it e.g.}, for the
(non-radiative) jet cross sections in DIS \cite{similar}), the phase
space slicing parameter $y_0^J$ has to be chosen very small, of the
order of $10^{-3}$ or smaller, in order to allow for the neglect of
terms of order $O(y_0^J)$. Therefore, $y_0^J$ can not be identified
with the $y$-cut of a jet algorithm applied in an experimental
analysis.  There, due to experimental restrictions, $y$ cannot be
reduced to values below $O(10^{-3})$. In addition, a fixed-order
calculation may give unphysical, {\it i.e.}\ negative $(1+1)$-jet
cross sections for too small values of $y$ (see the curves labeled
with S in Figs.\ \ref{fig2}a and \ref{fig3}a below). The Monte Carlo
approach, however, allows to apply a jet algorithm to the parton-level
events, {\it i.e.}\ to recombine 2 partons in the parton-level
$(2+1)$-jet events according to a jet algorithm using $y$-cuts $y^J$
for the separation of jet pairs (similarly: $y^{\gamma}$ for the
separation of a jet and a photon) with values as appropriate for the
given experimental situation.

The calculation thus proceeds through two subsequent steps: First, phase
space slicing is applied with a small $y$-cut $y_0^J$ of the order of
$\lsim 10^{-3}$ to accomplish the cancellation of singularities.  This
step relies on analytic calculations. Secondly, a jet algorithm is
applied with experimentally realizable, {\it i.e.}\ large enough values
$y^J$ and $y^{\gamma}$ of the order of $0.01 - 0.1$. The second step is
performed during the Monte Carlo integration.

The singular contributions for the process $eq \rightarrow eqg\gamma$
involve the following factors 
\begin{equation}
\left\{ \frac{1}{y_{36}}, \frac{1}{y_{46}}\right\}, 
\end{equation}
those for $eg \rightarrow eq\bar{q}\gamma$ contain the factors
\begin{equation}
\left\{ \frac{1}{y_{36}}, \frac{1}{y_{34}}\right\}.
\end{equation}
Terms containing $1/y_{3i}$, {\it i.e.}\ the momentum $p_3$ of the
incoming parton, are associated to initial-state singularities, terms
that do not, to final-state singularities. Contributions involving the
product of an ISR and an FSR factor, as for example the factor
$1/y_{36}y_{46}$, can be separated by partial fractioning,
\begin{equation}
\frac{1}{y_{ij}y_{ik}} = \frac{1}{y_{ij}} \frac{1}{y_{ij}+y_{ik}}
                       + \frac{1}{y_{ik}} \frac{1}{y_{ij}+y_{ik}},
\end{equation}
so that all singular contributions can be associated either to the
initial state or to the final state. Note that the denominator
$y_{ij}+y_{ik}$ introduced by partial fractioning can become zero only
if both $y_{ij} = 0$ and $y_{ik} = 0$ at the same time; since
configurations where all three partons 3, 4, and 6 are collinear with
each other are excluded by the cut on $W$, this is possible only for
$p_i = 0$. Therefore, for a contribution containing the pole factor
$1/y_{ij}$, we can separate the phase space into three regions:
\begin{itemize}
\item $y_{ij} < y_0^J$. This region contains the infrared singularity
  at $y_{ij} = y_{ik} = 0$, as well as the collinear singularity at
  $y_{ij} = 0$, $y_{ik} > 0$ and leads to singular contributions, {\it
    i.e.}\ $1/\epsilon$ and $1/\epsilon^2$ poles in dimensional
  regularization.  The double-poles $1/\epsilon^2$ and parts of the
  single-poles $1/\epsilon$ cancel with corresponding singular
  contributions from virtual corrections. The remaining
  $1/\epsilon$-pole contributions are associated to the initial state,
  can be factorized, and are absorbed by renormalizing the parton
  distribution functions. The analytical integration over this phase
  space region is performed with the approximation of small $y_0^J$,
  {\it i.e.}\ neglecting terms of $O(y_0^J)$. This contribution will be
  denoted by ``S" (singular) below.
\item $y_{ij} \geq y_0^J$ and $y_{ik} \geq y_0^J$ with only parton-level
  \gjt\ events, denoted by ``R" (real corrections);
\item $y_{ij} \geq y_0^J$ and $y_{ik} < y_0^J$. Here, the result is
  non-singular (therefore denoted by ``F", finite) but does not vanish
  with $y_0^J \rightarrow 0$, contrary to naive expectations. Its 
  contribution is calculated numerically. It is non-negligible in
  particular for terms related to ISR singularities.
\end{itemize}

The integrals needed for the singular contributions are written in a
Lorentz-invariant form as tensor integrals which can be reduced to a
few basic scalar integrals with the help of analytic programs like
{\tt mathematica} or {\tt form}. More details are given in
\cite{michelsen}.  The remaining phase space integrations are
performed with the help of Monte Carlo techniques. The three
contributions S, R, and F are treated separately, each with
appropriate mappings of the respective integration variables to
improve the numerical stability of the calculation.

As discussed in the introduction, in the present work we do not
factorize and subtract those photon-parton collinear singularities
which have to be absorbed into parton-to-photon fragmentation
functions. Instead, we remove all singular contributions by keeping
isolation cuts at the parton level. As stated in the introduction,
care has to be taken that the isolation criteria do not restrict the
phase space for gluon emission since this would destroy the
cancellation of singular contributions. Therefore, in the first step
of the calculation described above, we require the photon to be
isolated from the quark (antiquark) by the cut
\begin{equation}
y_{5i} > y_0^{\gamma}
\label{pisocut}
\end{equation}
with $i = 3, 4$ for $eq \rightarrow eqg\gamma$ and $i = 4, 6$ for $eg
\rightarrow eq\bar{q}\gamma$. The cut is not applied to photon-gluon
pairs which is possible since gluons do not emit photons and there is
no singularity related to $y_{g\gamma}$. This definition of photon
isolation at the parton level introduces an unphysical parameter
($y_0^{\gamma}$). The sensitivity to $y_0^{\gamma}$ can be reduced by
applying, in the second step of the calculation, photon isolation with
respect to jets described by cutoff parameters which can be used in
the same way in the experimental analysis. In order to have some
freedom when modeling these physical isolation criteria we choose a
small value for $y_0^{\gamma}$. The dependence on $y_0^{\gamma}$ will
be discussed below. 
 

\section{Numerical Results}

The results discussed in the following are obtained for energies and
cuts appropriate for the HERA experiments: the energies of the
incoming electron (positron) and proton are $E_e = 27.5$ GeV, $E_P =
820$ GeV and
\begin{equation}
\begin{array}{c}
Q^2 \geq 10~{\rm GeV}^2, ~~~~
W   \geq 10~{\rm GeV}  , ~~~~ \\[1ex]
0.001 \leq x \leq 0.5, ~~~~ 0.05 \leq y \leq 0.99, ~~~~ \\[1ex]
p^T_{\gamma} \geq 5~{\rm GeV}, ~~~~ 
90^{\circ} \leq \theta_{\gamma} \leq 170^{\circ}, ~~~~
\theta_{\gamma e} \geq 10^{\circ}.
\end{array}
\label{allcuts}
\end{equation}
Note that the emission angle of the photon, $\theta_{\gamma}$,
measured with respect to the incoming electron in the HERA laboratory
frame, is restricted to the hemisphere $\theta_{\gamma} \geq
90^{\circ}$ since photon production with $\theta_{\gamma} <
90^{\circ}$ is dominated by `uninteresting' leptonic radiation.
The parton distribution functions are taken from Ref.\ \cite{MRSA} 
(MRS(A)).  

The events generated during Monte Carlo integration are $\gamma q$,
$\gamma q g$ or $\gamma q \bar{q}$ events.  A simple event analysis is
applied to obtain \gjo\ and \gjt\ event samples. The event analysis
consists of two parts: the first part serves to identify the number of
jets according to a conventional jet algorithm; the second part treats
photon isolation. For simplicity we choose a jet definition using the
normalized invariant masses $y_{ij}$. Since for small $\xi \ge x$ the
momentum of the remnant and thus $y_{ir}$ can be large even for
partons with small transverse momentum, we first remove low-$p^T$
partons before recombining partons to jets. Explicitly we apply the
following conditions: 
\begin{itemize}
\item[(1)] A final state parton (quark or gluon) is recombined with the
  remnant if its transverse momentum is below a cutoff:
  \begin{equation}
  p^T_i < p^T_{\rm min} = 1~{\rm GeV}, ~~~~ i=4,6.
  \label{ptmincut}
  \end{equation}
\item[(2)] Two partons are recombined into one jet if
  \begin{equation}
  y_{ij} < y^J ~~~~ {\rm for} ~~~~ i, j = 4, 6, r
  \end{equation}
  and all quarks, antiquarks and gluons as well as the proton remnant
  are taken into account when forming jets. If several pairs of
  partons have $y_{ij}$ below $y^J$, the pair with the smallest
  $y_{ij}$ is recombined first. Several prescriptions for the
  recombination are possible: the energy and 3-momentum of a jet
  $(ij)$ obtained from pairing partons $i$ and $j$ can be obtained by
  \begin{equation}
  E_{ij} = \alpha (E_i + E_j), ~~~~
  \vec{p}_{ij} = \beta (\vec{p}_i + \vec{p}_j).
  \end{equation}
  For example in the E-scheme one chooses $\alpha = \beta = 1$; in the
  P-scheme one has $\alpha = \left| \vec{p}_i + \vec{p}_j \right| /
  (E_i + E_j)$, $\beta = 1$ instead.  Since the present calculation is
  of first order in $\alpha_s$, the recombination has not to be
  iterated. However, the different recombination prescriptions become
  relevant when photon isolation with respect to jets is imposed. In
  addition, the cut on low-$p_T$ partons or parton-pairs Eq.\ 
  (\ref{ptmincut}) is affected if a recombination prescription with
  $\beta \neq 1$ is used.  Our numerical results will be given for the
  P-scheme.
\item[(3)] Finally, an event is accepted only if the photon is
  separated from the jets or if the photon is accompanied by hadronic
  energy less than a specified amount, {\it i.e.}\ we exclude events
  with 
  \begin{equation}
  y_{\gamma j} < y^{\gamma} ~~~{\rm and}~~~
  E_j > \epsilon \left(E_j + E_{\gamma}\right)
  \label{epshad}
  \end{equation}
  where $j$ denotes any jet ({\it i.e.}, parton or pair of partons)
  remaining after steps (1) and (2) of the event analysis. 
\end{itemize}
We keep the possibility to use different values for the $y$-cuts applied
to purely hadronic jets and to jets containing the photon. In practice, 
$y^J$ and $y^{\gamma}$ are taken equal with a typical value 0.03. For
the photon isolation parameter $\epsilon$ we will take the value 0.1
as used in experimental analyses \cite{zeusph}. Apart from being
experimentally unrealistic, the value $\epsilon = 0$ is theoretically
not allowed since Eq.\ (\ref{epshad}) with $\epsilon = 0$ would
restrict the phase space for soft partons and consequently destroy the
cancellation of corresponding singularities. 

\begin{figure}[hptb] 
\unitlength 1mm
\begin{picture}(100,80)
\put(0,-1){\epsfxsize=8cm \epsfysize=8.5cm
\epsfbox{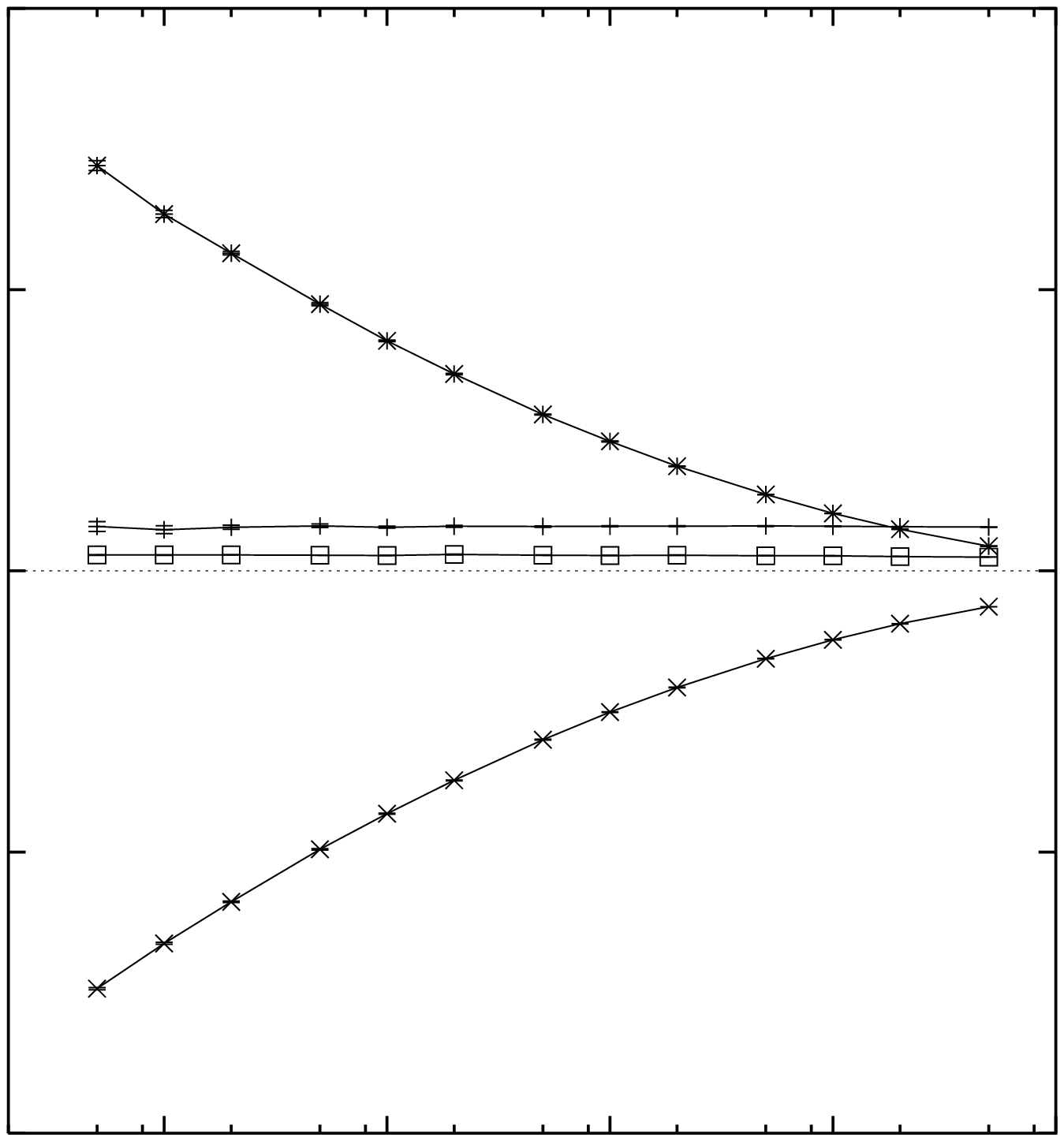}}
\put(6,82){\makebox(0,0)[r]{\small $100$}}
\put(6,63){\makebox(0,0)[r]{\small $50$}}
\put(6,43.8){\makebox(0,0)[r]{\small $0$}}
\put(6,24.5){\makebox(0,0)[r]{\small $-50$}}
\put(6,5.5){\makebox(0,0)[r]{\small $-100$}}
\put(14,1){\small $10^{-6}$}
\put(29,1){\small $10^{-5}$}
\put(44,1){\small $10^{-4}$}
\put(59,1){\small $10^{-3}$}
\put(74,1){\small $10^{-2}$}
\put(68,0){$y_0^J$}
\put(16,70){R}
\put(16,40){F}
\put(16,13){S}
\put(16,48.5){sum (incl. LO)}
\put(30,73){$\sigma(\gamma+(1+1)$-jets) [pb]}
\put(30,68){incoming $q$, $\bar{q}$}
\put(40,-3){a)}
\put(83,-1){ \epsfxsize=8cm \epsfysize=8.5cm
\epsfbox{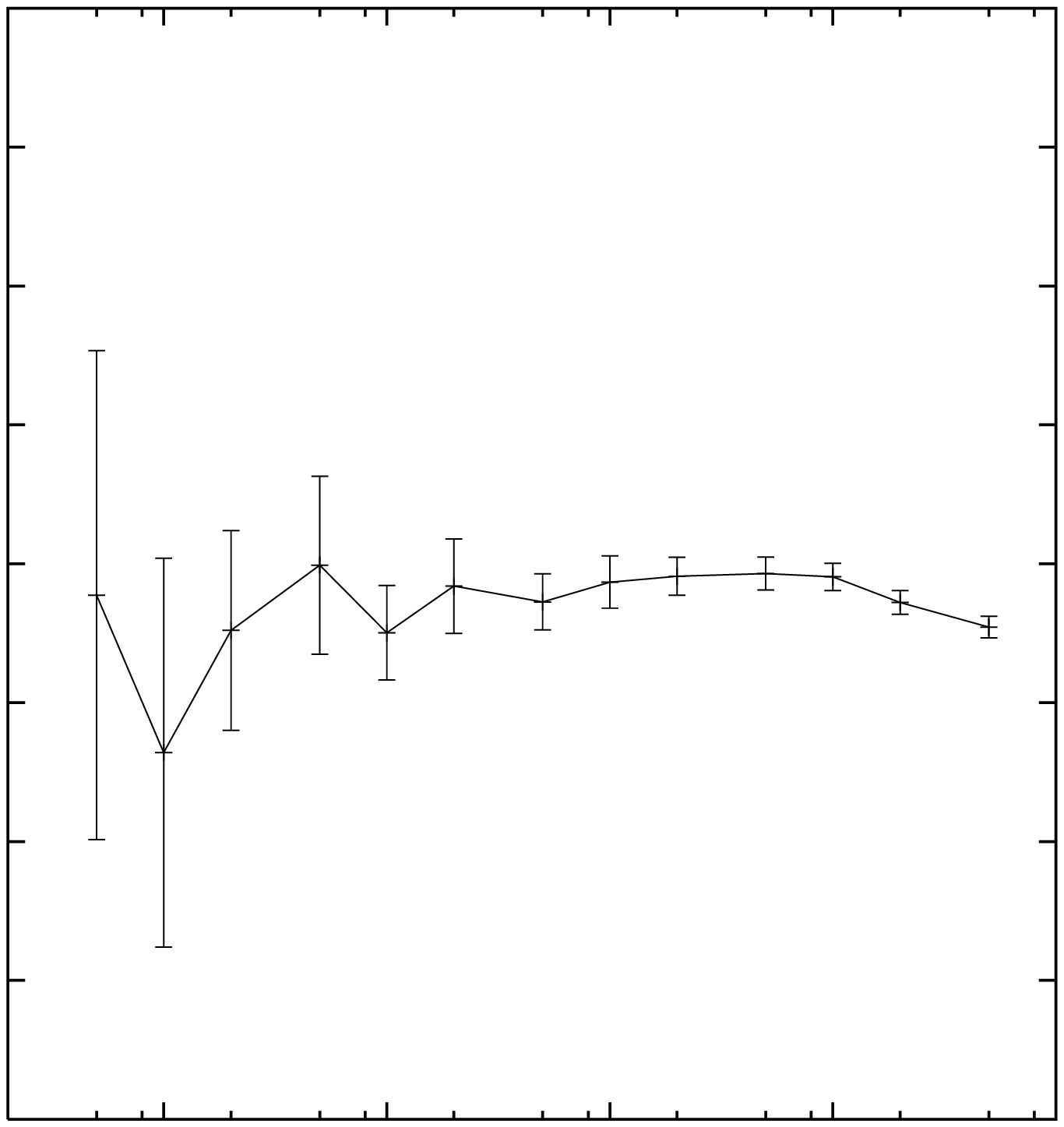}}
\put(89,82){\makebox(0,0)[r]{\small $10$}}
\put(89,72.5){\makebox(0,0)[r]{\small $9.5$}}
\put(89,63){\makebox(0,0)[r]{\small $9$}}
\put(89,53.5){\makebox(0,0)[r]{\small $8.5$}}
\put(89,44){\makebox(0,0)[r]{\small $8$}}
\put(89,34.5){\makebox(0,0)[r]{\small $7.5$}}
\put(89,25){\makebox(0,0)[r]{\small $7$}}
\put(89,15.5){\makebox(0,0)[r]{\small $6.5$}}
\put(89,6){\makebox(0,0)[r]{\small $6$}}
\put(98,1){\small $10^{-6}$}
\put(113,1){\small $10^{-5}$}
\put(128,1){\small $10^{-4}$}
\put(143,1){\small $10^{-3}$}
\put(158,1){\small $10^{-2}$}
\put(152,0){$y_0^J$}
\put(105,73){$\sigma(\gamma+(1+1)$-jets) [pb]}
\put(105,68){incoming $q$, $\bar{q}$}
\put(208,-3){b)}
\end{picture}
\caption{\it Dependence of the \gjo\ cross section on the phase
  space slicing cut $y_0^J$ for incoming quarks and
  antiquarks ($y_0^{\gamma}=10^{-4}$, $y^{\gamma} = y^J = 0.03$). (a)
  shows the separate contributions and the sum = R + S + F + LO. (b)
  shows the sum of all contributions, including the leading-order
  cross section, on a larger scale.} 
\label{fig2}
\end{figure}

\begin{figure}[hpbt] 
\unitlength 1mm
\begin{picture}(100,80)
\put(2,-1){\epsfxsize=7.8cm \epsfysize=8.5cm
\epsfbox{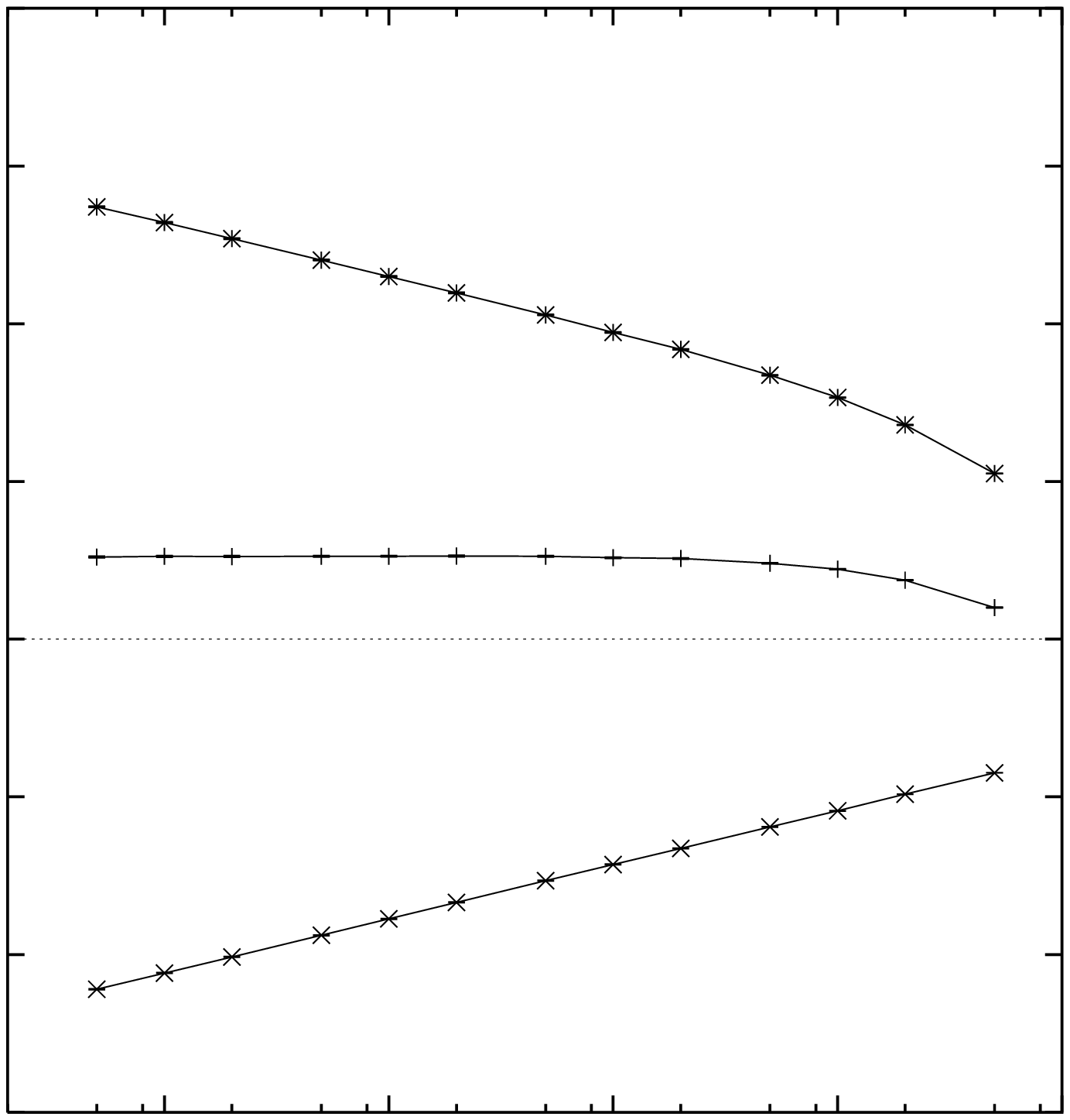}}
\put(14,1){\small $10^{-6}$}
\put(29,1){\small $10^{-5}$}
\put(44,1){\small $10^{-4}$}
\put(59,1){\small $10^{-3}$}
\put(74,1){\small $10^{-2}$}
\put(68,0){$y_0^J$}
\put(6,81.9){\makebox(0,0)[r]{\small $8$}}
\put(6,71.2){\makebox(0,0)[r]{\small $6$}}
\put(6,60){\makebox(0,0)[r]{\small $4$}}
\put(6,49.1){\makebox(0,0)[r]{\small $2$}}
\put(6,38.3){\makebox(0,0)[r]{\small $0$}}
\put(6,27.4){\makebox(0,0)[r]{\small $-2$}}
\put(6,16.6){\makebox(0,0)[r]{\small $-4$}}
\put(6,5.8){\makebox(0,0)[r]{\small $-6$}}
\put(16,63){R}
\put(16,17){S}
\put(16,45.5){sum}
\put(30,74){$\sigma(\gamma+(1+1)$-jets) [pb]}
\put(30,69){incoming $g$}
\put(40,-3){a)}
\put(83,-1){ \epsfxsize=8cm \epsfysize=8.5cm
\epsfbox{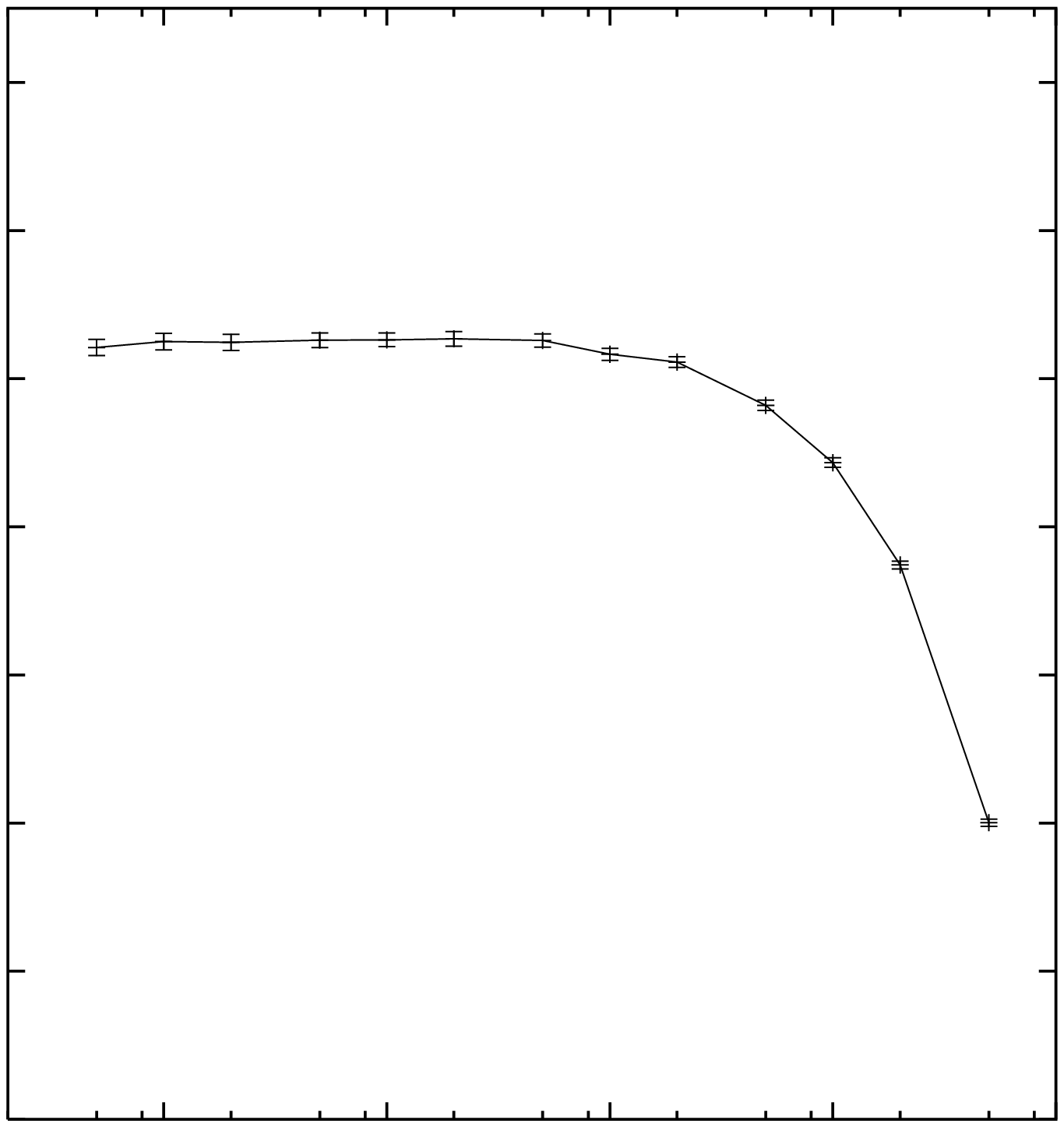}}
\put(98,1){\small $10^{-6}$}
\put(113,1){\small $10^{-5}$}
\put(128,1){\small $10^{-4}$}
\put(143,1){\small $10^{-3}$}
\put(158,1){\small $10^{-2}$}
\put(152,0){$y_0^J$}
\put(89,76.7){\makebox(0,0)[r]{\small $1.4$}}
\put(89,66.6){\makebox(0,0)[r]{\small $1.2$}}
\put(89,56.5){\makebox(0,0)[r]{\small $1$}}
\put(89,46.4){\makebox(0,0)[r]{\small $0.8$}}
\put(89,36.3){\makebox(0,0)[r]{\small $0.6$}}
\put(89,26.2){\makebox(0,0)[r]{\small $0.4$}}
\put(89,16.1){\makebox(0,0)[r]{\small $0.2$}}
\put(89,6){\makebox(0,0)[r]{\small $0$}}
\put(105,28){$\sigma(\gamma+(1+1)$-jets) [pb]}
\put(105,23){incoming $g$}
\put(124,-3){b)}
\end{picture}
\caption{\it Dependence of the \gjo\ cross section on the phase
  space slicing cut $y_0^J$ for incoming gluons ($y_0^{\gamma}=10^{-4}$,
  $y^{\gamma} = y^J = 0.03$). (a) shows the separate contributions and
  the sum = R + S. (b) shows the sum of all contributions on a larger
  scale.}
\label{fig3}
\end{figure}

We start with demonstrating the consistency of our approach by showing
the dependence of the \gjo\ cross section on the phase space slicing
cut $y_0^J$. Figures \ref{fig2} and \ref{fig3} show the dependence of
the total and partial cross sections for \gjo\ events. For
$q(\bar{q})$-initiated processes, the separate contributions S and R
depend on $\log^2 y_0^J$ (see Fig.\ \ref{fig2}a) and the finite
contribution F is not negligible. In this case the sum is numerically
stable in the range $10^{-5} \lsim y_0^J \lsim 10^{-3}$; for smaller
values the numerical precision decreases and for larger values the
error from neglected terms of order $O(y_0^J)$ is not negligible. The
calculation of $g$-initiated contributions can be performed with much
smaller uncertainties and for much smaller values of $y_0^J$, as seen
in Fig.\ \ref{fig3} since here the dependence on $\log y_0^J$ is only
linear.  Also, the finite contribution is negligible for diagrams with
incoming gluons. The dependence on $y_0^J$ at large values above
$\simeq 10^{-3}$ is slightly stronger in this case than for
$q(\bar{q})$-initiated contributions since terms of order $O(y_0^J)$
are relatively more important. In the following we fix $y_0^J$ at the
value $10^{-4}$. 

\begin{figure}[hpbt] 
\unitlength 1mm
\begin{picture}(100,80)
\put(1,-1){\epsfxsize=8cm \epsfysize=8.5cm
\epsfbox{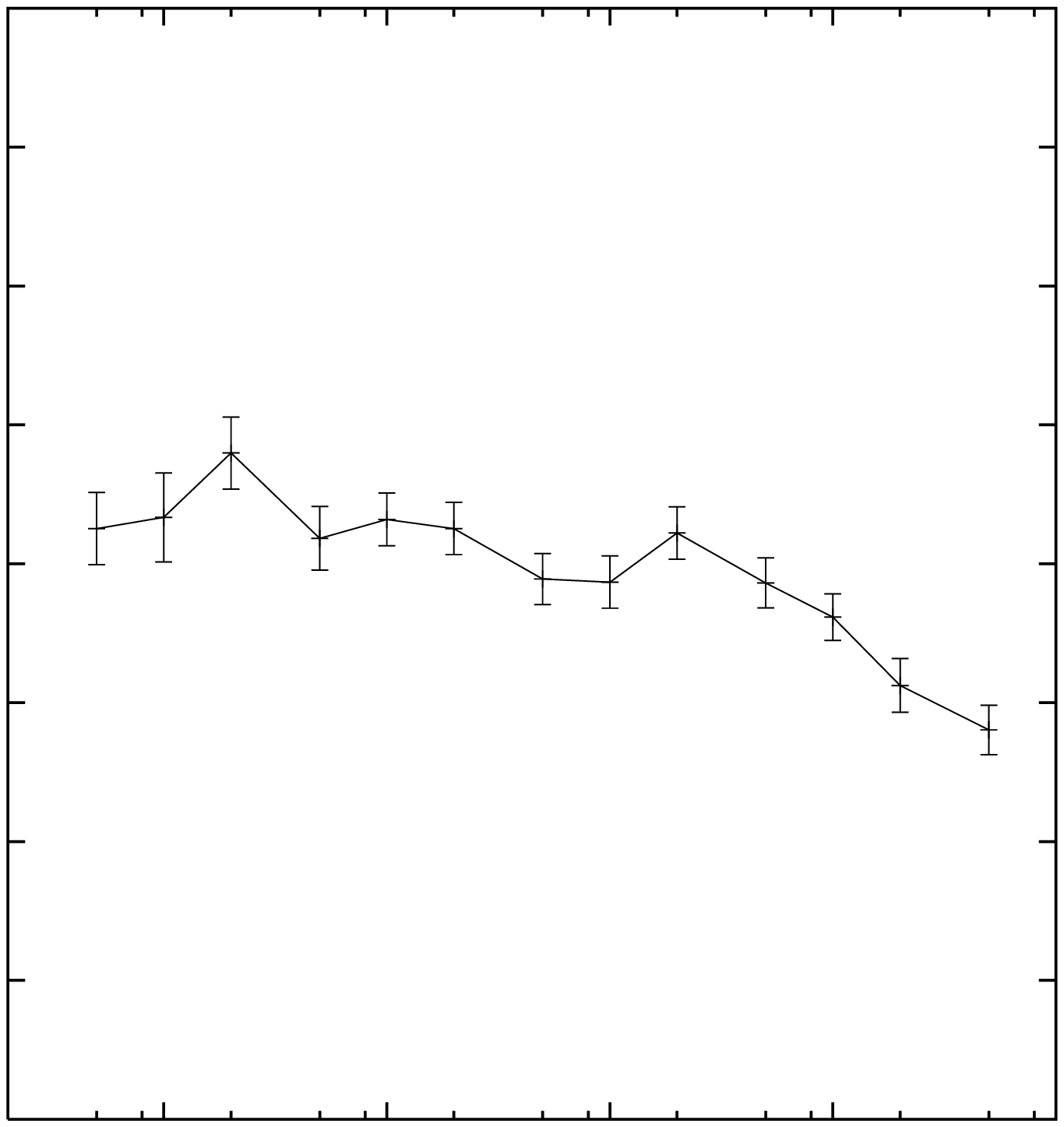}}
\put(14.2,1){\small $10^{-6}$}
\put(29.4,1){\small $10^{-5}$}
\put(44.6,1){\small $10^{-4}$}
\put(59.8,1){\small $10^{-3}$}
\put(69,0){$y_0^{\gamma}$}
\put(76,1){\small $10^{-2}$}
\put(6,82){\makebox(0,0)[r]{\small $10$}}
\put(6,72.5){\makebox(0,0)[r]{\small $9.5$}}
\put(6,63){\makebox(0,0)[r]{\small $9$}}
\put(6,53.5){\makebox(0,0)[r]{\small $8.5$}}
\put(6,44){\makebox(0,0)[r]{\small $8$}}
\put(6,34.5){\makebox(0,0)[r]{\small $7.5$}}
\put(6,25){\makebox(0,0)[r]{\small $7$}}
\put(6,15.5){\makebox(0,0)[r]{\small $6.5$}}
\put(6,6){\makebox(0,0)[r]{\small $6$}}
\put(16,73){$\sigma(\gamma+(1+1)$-jets) [pb]}
\put(16,68){incoming $q$, $\bar{q}$}
\put(34,-3){a)}
\put(83,-1){ \epsfxsize=8cm \epsfysize=8.5cm
\epsfbox{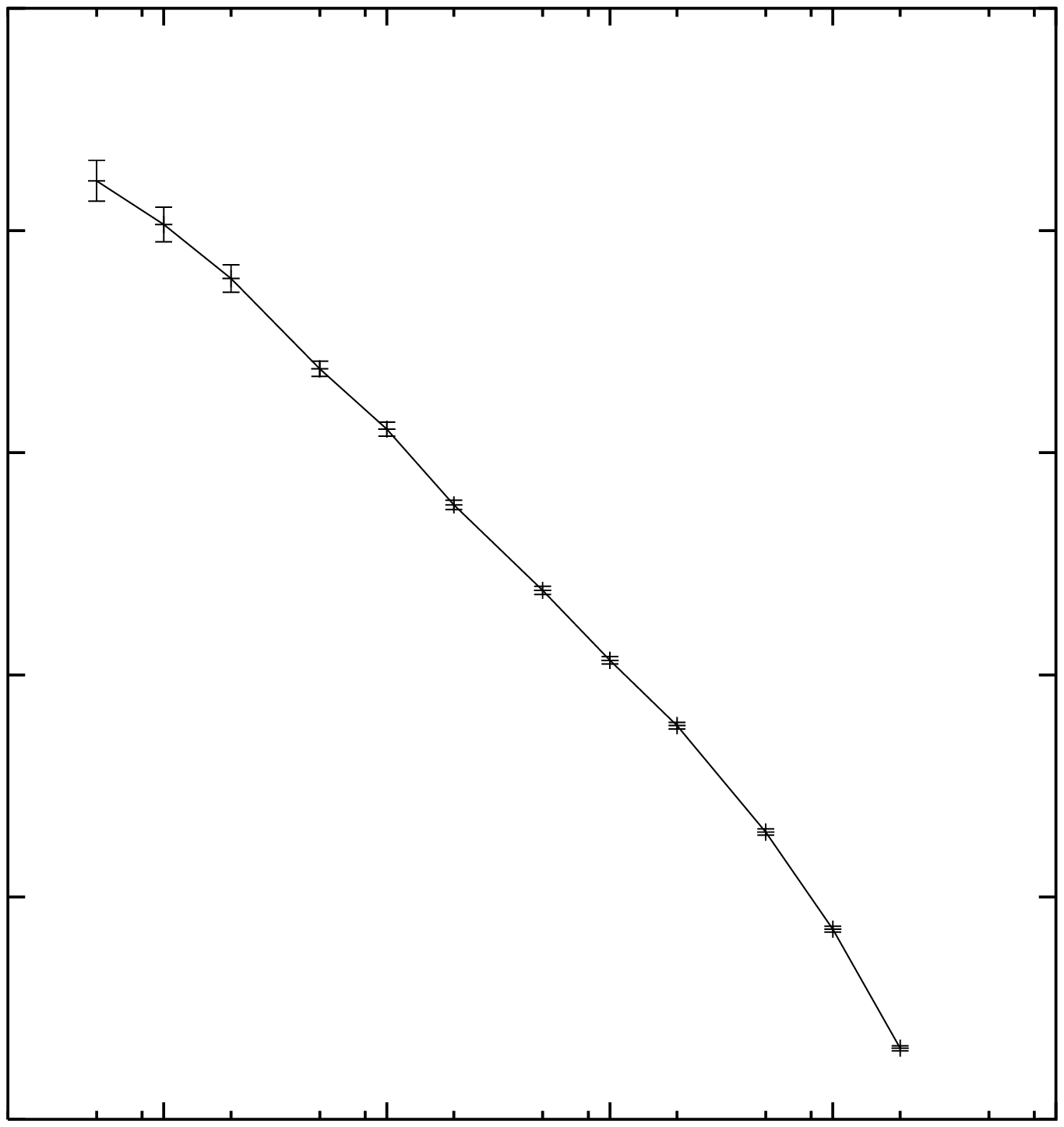}}
\put(98,1){\small $10^{-6}$}
\put(113,1){\small $10^{-5}$}
\put(128,1){\small $10^{-4}$}
\put(143,1){\small $10^{-3}$}
\put(158,1){\small $10^{-2}$}
\put(152,0){$y_0^{\gamma}$}
\put(89,82){\makebox(0,0)[r]{\small $2.5$}}
\put(89,66.8){\makebox(0,0)[r]{\small $2$}}
\put(89,51.6){\makebox(0,0)[r]{\small $1.5$}}
\put(89,36.4){\makebox(0,0)[r]{\small $1$}}
\put(89,21.2){\makebox(0,0)[r]{\small $0.5$}}
\put(89,6){\makebox(0,0)[r]{\small $0$}}
\put(113,73){$\sigma(\gamma+(1+1)$-jets) [pb]}
\put(113,68){incoming $g$}
\put(124,-3){b)}
\end{picture}
\caption{\it Dependence on the infrared cutoff parameter
  $y_0^{\gamma}$ of the \gjo\ cross section for incoming
  (anti)quarks (a) and gluons (b) ($y_0^J=10^{-4}$, $y^{\gamma} =
  y^J = 0.03$).}
\label{fig4}
\end{figure}

\begin{figure}[htb] 
\unitlength 1mm
\begin{picture}(100,105)
\put(33,-17){\epsfxsize=10cm \epsfysize=13cm
\epsfbox{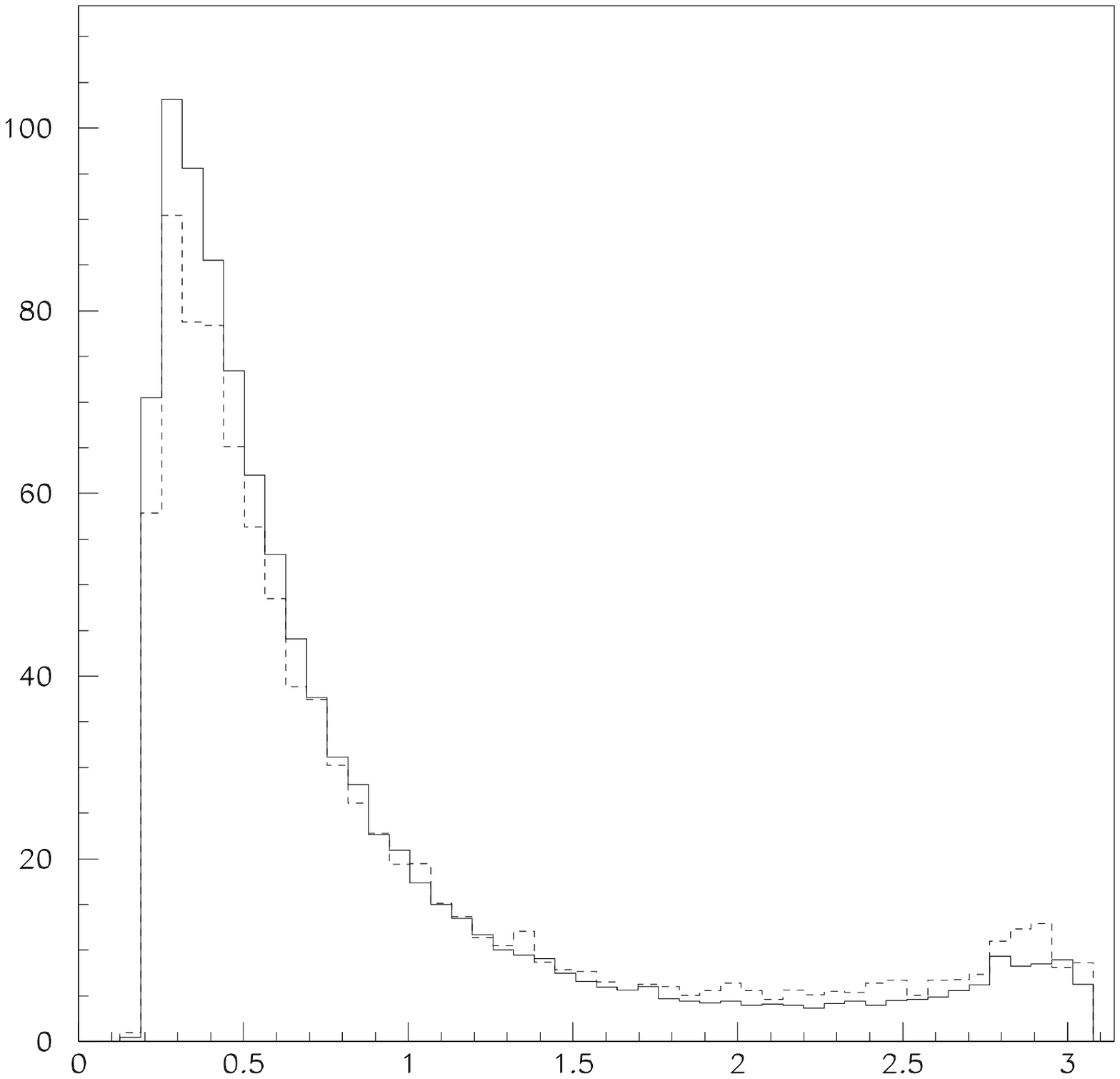}}
\put(112,1){$\theta_{\gamma} [{\rm rad}]$}
\put(100,83){$\displaystyle \frac{d\sigma}{d\theta_{\gamma}}$ [pb]} 
\put(46,88){Born}
\put(114,23){${\cal O}(\alpha_s)$}
\end{picture}
\caption{\it Differential cross section $d\sigma/d\theta_{\gamma}$ for
  $ep \rightarrow e\gamma + (1+1)$-jets and $ep \rightarrow e\gamma +
  (2+1)$-jets with $y^J = y^{\gamma} = 0.03$. Full histogram: lowest
  order, dashed histogram: including $O(\alpha_s)$ corrections. Cuts are
  explained in the text.}
\label{fig5}
\end{figure}

Figure \ref{fig4} shows the residual dependence of the cross sections
for \gjo s on the parton-level photon isolation cut $y_0^{\gamma}$,
separately for (anti)quark and gluon-initiated processes. The
$y_0^{\gamma}$-dependence is weak for the case with incoming
(anti)quarks showing that the isolation criteria efficiently reduce
the sensitivity to the phase space region where the non-perturbative
parton-to-photon fragmentation functions would contribute. For the
$g$-initiated processes, the sensitivity to $y_0^{\gamma}$ is larger.
In this case all final-state partons ($q$ and $\bar{q}$) can emit a
photon. Since the isolation condition is applied to jets, the
singularity associated to configurations with soft (anti)quarks having
emitted a hard collinear photon is removed only with the help of the
parton-level cut Eq.\ (\ref{pisocut}). For processes with incoming
quarks, only a small subset of diagrams leads to singularities for
similar configurations.  We choose $y_0^{\gamma} = 10^{-4}$ in the
following. This value is small enough compared with experimentally
realistic values for $y^{\gamma} \gsim O(10^{-2})$ so that
contributions where a quark or an antiquark determines the momentum of
a jet, become insensitive to $y_0^{\gamma}$. Also, much larger values
would lead to an unphysical negative cross section for the
$g$-initiated subprocess.  In a more systematic treatment, the
$y_0^{\gamma}$-dependent terms in our calculation would be replaced by
contributions from parton-to-photon fragmentation functions.  In our
present approach, however, the unwanted dependence on $y_0^{\gamma}$
has to be viewed as an unavoidable source of a theoretical
uncertainty. The \gjo cross section at $Q^2 \lsim 100$ GeV${}^2$ is
affected by this at the level of 20\,\% (see Fig.\ \ref{fig6} below).
At larger $Q^2$, the influence of the $y_0^{\gamma}$-dependent
gluon-initiated contribution is reduced\footnote{This can be compared
  with the case of $e^+e^- \rightarrow \gamma +$hadrons where the
  total cross section has little sensitivity to the parton-level
  photon isolation cut for not too large $y$, but the $\gamma + 1$-jet
  rate has a non-negligible dependence on $y_0^{\gamma}$ \cite{msz}.}.

In Fig.\ \ref{fig5} we show the differential cross section
$d\sigma/d\theta_{\gamma}$ (sum of \gjo s and \gjt s) in the range
$10^{\circ} \leq \theta_{\gamma} \leq 175^{\circ}$. Apart from the
extended range of photon emission angles all cuts given in
(\ref{allcuts}) are applied.  The majority of photons is produced with
small angles, {\it i.e.}\ close to the direction of the incoming lepton.
For leptonic radiation, QCD corrections reduce the cross section by
$\sim 10\,\%$ for the phase space region under consideration. By
contrast, at large emission angles, dominated by quarkonic radiation,
the cross section receives positive QCD corrections. In the following we
restrict ourselves again to this ``signal" region $\theta_{\gamma} \ge
90^{\circ}$, {\it i.e.}\ the proton hemisphere in the HERA laboratory
system.

\begin{figure}[htb] 
\unitlength 1mm
\begin{picture}(100,105)
\put(33,-17){\epsfxsize=10cm \epsfysize=13cm
\epsfbox{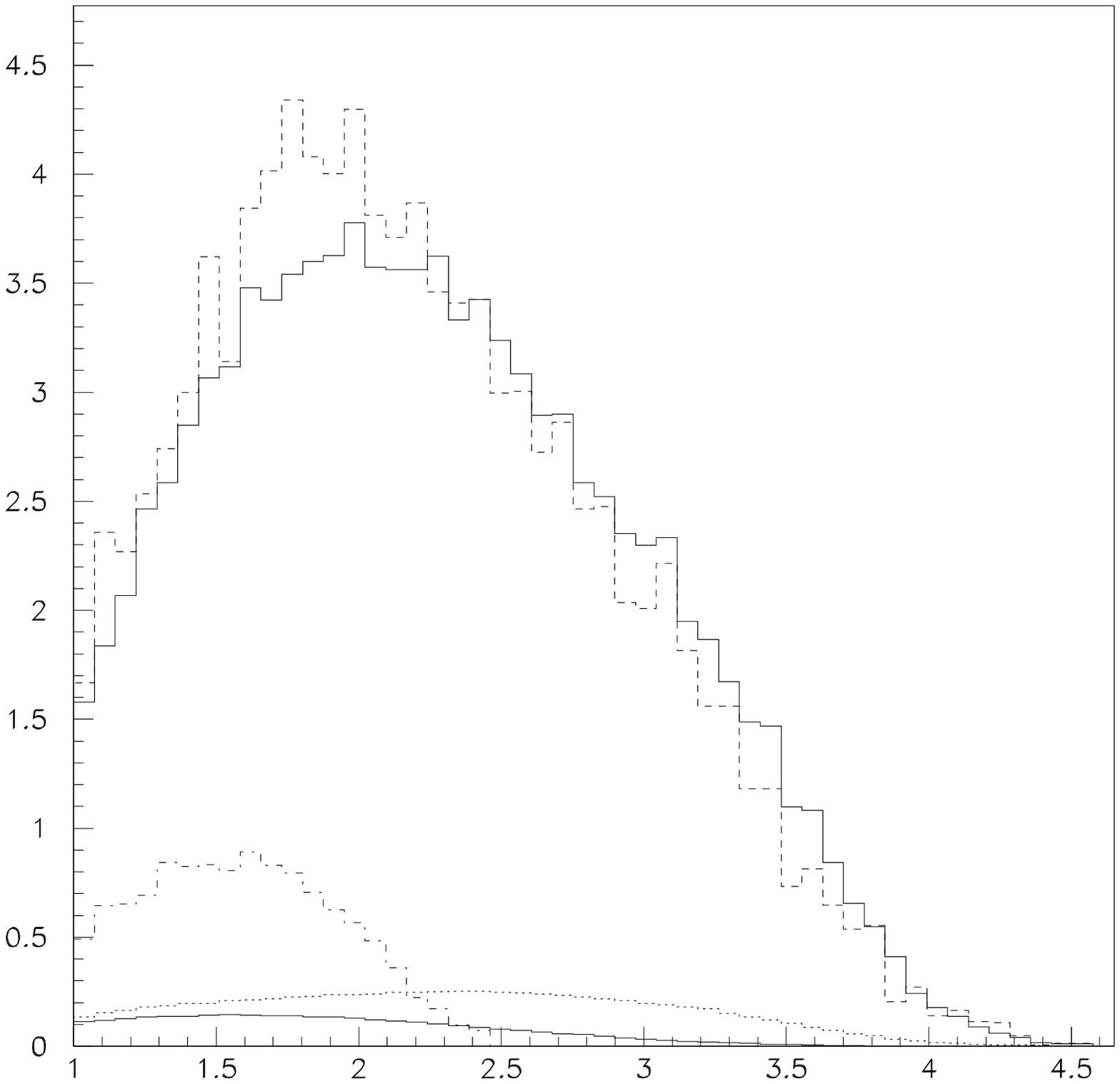}}
\put(99,0){$\log_{10}(Q^2/{\rm GeV}^2)$}
\put(100,85){$\displaystyle Q^2\frac{d\sigma}{dQ^2}$ [pb]}
\put(96,75){$90^{\circ}\le\theta_{\gamma}\le 170^{\circ}$}
\put(42,50){Born (1+1)}
\put(40,92){$q+\bar{q}$ (1+1)}
\put(76,15){$q+\bar{q}$ (2+1)}
\put(50,27){$g$ (1+1)}
\put(45,15.5){$g$ (2+1)}
\put(46,14){\line(2,-1){6}}
\end{picture}
\caption{\it Differential cross section $Q^2 d\sigma/dQ^2$ for $ep
  \rightarrow e\gamma + (n+1)$-jets with $y^J = y^{\gamma} = 0.03$.
  Upper full histogram: lowest order, dashed histogram: contribution
  from incoming (anti)quarks to \gjo s, dotted histogram: contribution
  from incoming (anti)quarks to \gjt s, dash-dotted histogram: incoming
  gluons for \gjo s, lower full histogram: incoming gluons for \gjt s.
  Photons are restricted to $90^{\circ}\le\theta_{\gamma}\le
  170^{\circ}$ and other cuts are explained in the text.}
\label{fig6}
\end{figure}

The $Q^2$-dependence in this restricted phase space region is shown in
Fig.\ \ref{fig6}. The cross section is shown separately for
$q(\bar{q})$-initiated and gluon-initiated contributions giving rise to
\gjo\ and \gjt\ events using $y^J = y^{\gamma} = 0.03$. The \gjo\ 
contribution is dominant for these $y$-cut values with a maximum in
the lower $Q^2$ range, whereas the cross section for \gjt\ events is
flatter and extends to larger $Q^2$. Incoming gluons contribute only
roughly 10\,\% to the total cross section. Since the distributions for
incoming quarks and incoming gluons are not very different, it seems
difficult to utilize radiative deep inelastic scattering for a
measurement of the gluon distribution. We also checked that using
other parametrizations of parton distribution functions (like those of
Refs.\ \cite{grvcteq}) do not lead to significantly different shapes
of distributions. Only the total cross sections vary by $O(10 -
15\,\%)$. 

\begin{figure}[hptb] 
\unitlength 1mm
\begin{picture}(100,100)
\put(33,-2){\epsfxsize=9.5cm \epsfysize=10.5cm
\epsfbox{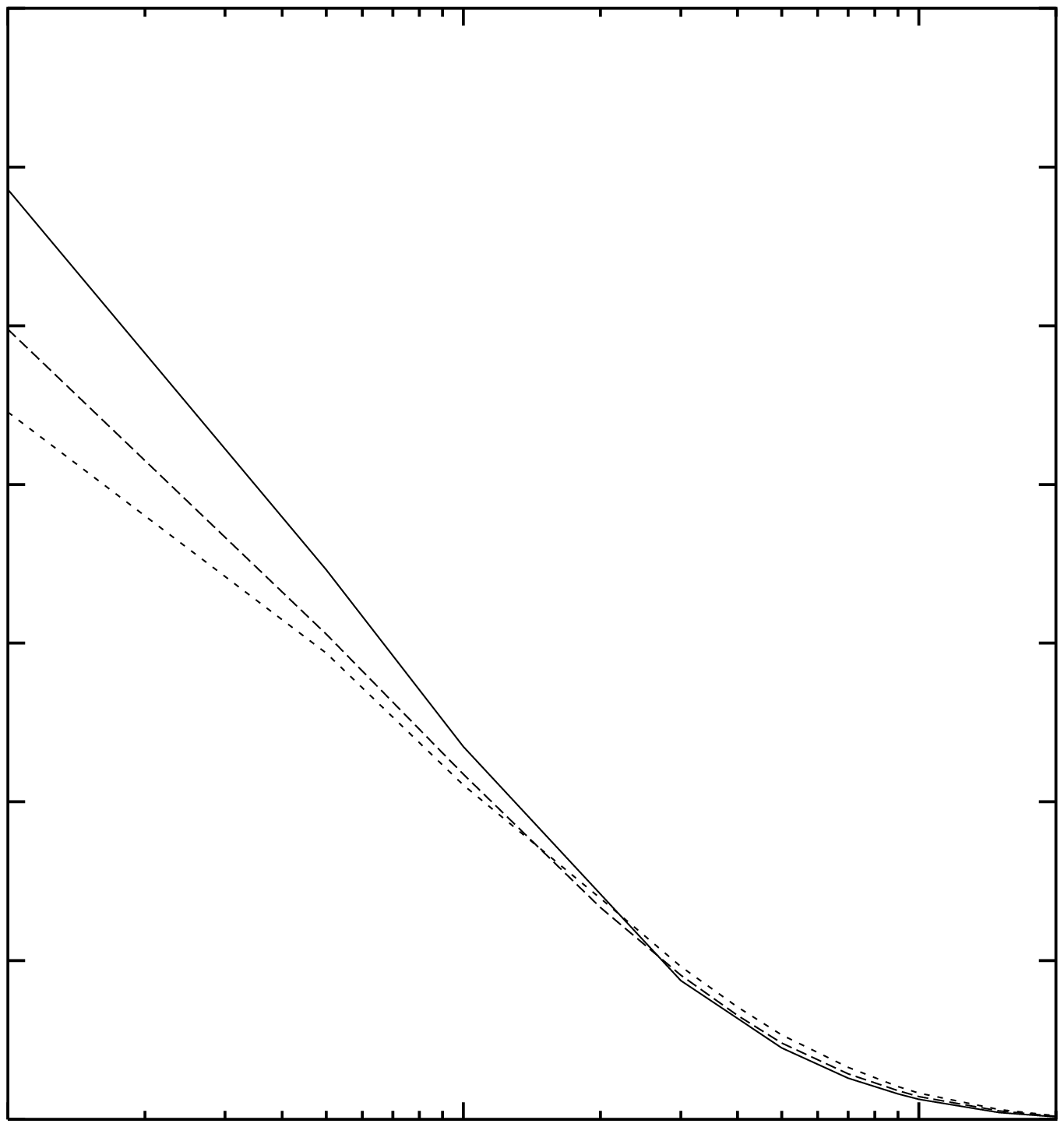}}
\put(38,1){\small $10^{-3}$}
\put(74,1){\small $10^{-2}$}
\put(111,1){\small $10^{-1}$}
\put(100,0){$y^J$}
\put(39,101){\makebox(0,0)[r]{\small $0.7$}}
\put(39,87.6){\makebox(0,0)[r]{\small $0.6$}}
\put(39,73){\makebox(0,0)[r]{\small $0.5$}}
\put(39,59.7){\makebox(0,0)[r]{\small $0.4$}}
\put(39,46.2){\makebox(0,0)[r]{\small $0.3$}}
\put(39,32.8){\makebox(0,0)[r]{\small $0.2$}}
\put(39,19.4){\makebox(0,0)[r]{\small $0.1$}}
\put(39,6){\makebox(0,0)[r]{\small $0$}}
\put(43,88){$y^{\gamma} = $}
\put(43,82){0.01}
\put(43,69.8){0.05}
\put(43,57){0.1}
\put(100,73){\large $R_{\gamma,2+1}$}
\end{picture}
\caption{\it \gjt\ rate $R_{\gamma,2+1}$ as a function of the jet cut
  $y^J$ for three different values of the photon isolation cut
  $y^{\gamma}$.} 
\label{fig7}
\end{figure}
\begin{figure}[hpbt] 
\unitlength 1mm
\begin{picture}(100,80)
\put(2,-1){\epsfxsize=8cm \epsfysize=8.5cm
\epsfbox{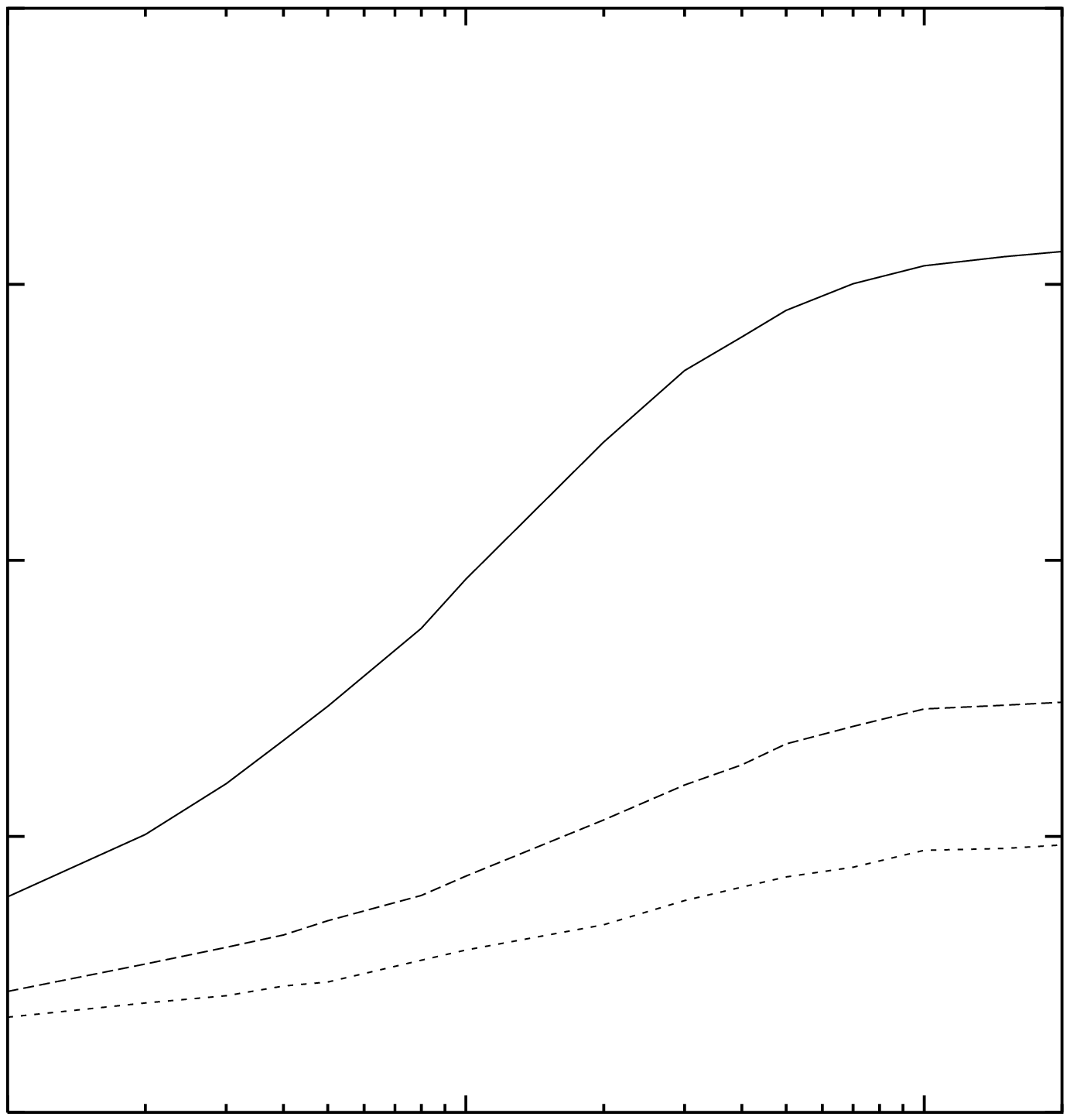}}
\put(4,1){\small $10^{-3}$}
\put(35.5,1){\small $10^{-2}$}
\put(67,1){\small $10^{-1}$}
\put(60,0){$y^J$}
\put(6,81.8){\makebox(0,0)[r]{\small $20$}}
\put(6,63.3){\makebox(0,0)[r]{\small $15$}}
\put(6,43.8){\makebox(0,0)[r]{\small $10$}}
\put(6,24.9){\makebox(0,0)[r]{\small $5$}}
\put(6,6){\makebox(0,0)[r]{\small $0$}}
\put(65,65){$y^{\gamma} = $}
\put(65,59){0.01}
\put(65,35){0.05}
\put(65,24.5){0.1}
\put(16,70){$\sigma(\gamma+(1+1)$-jets) [pb]}
\put(34,-3){a)}
\put(83,-1){ \epsfxsize=8cm \epsfysize=8.5cm
\epsfbox{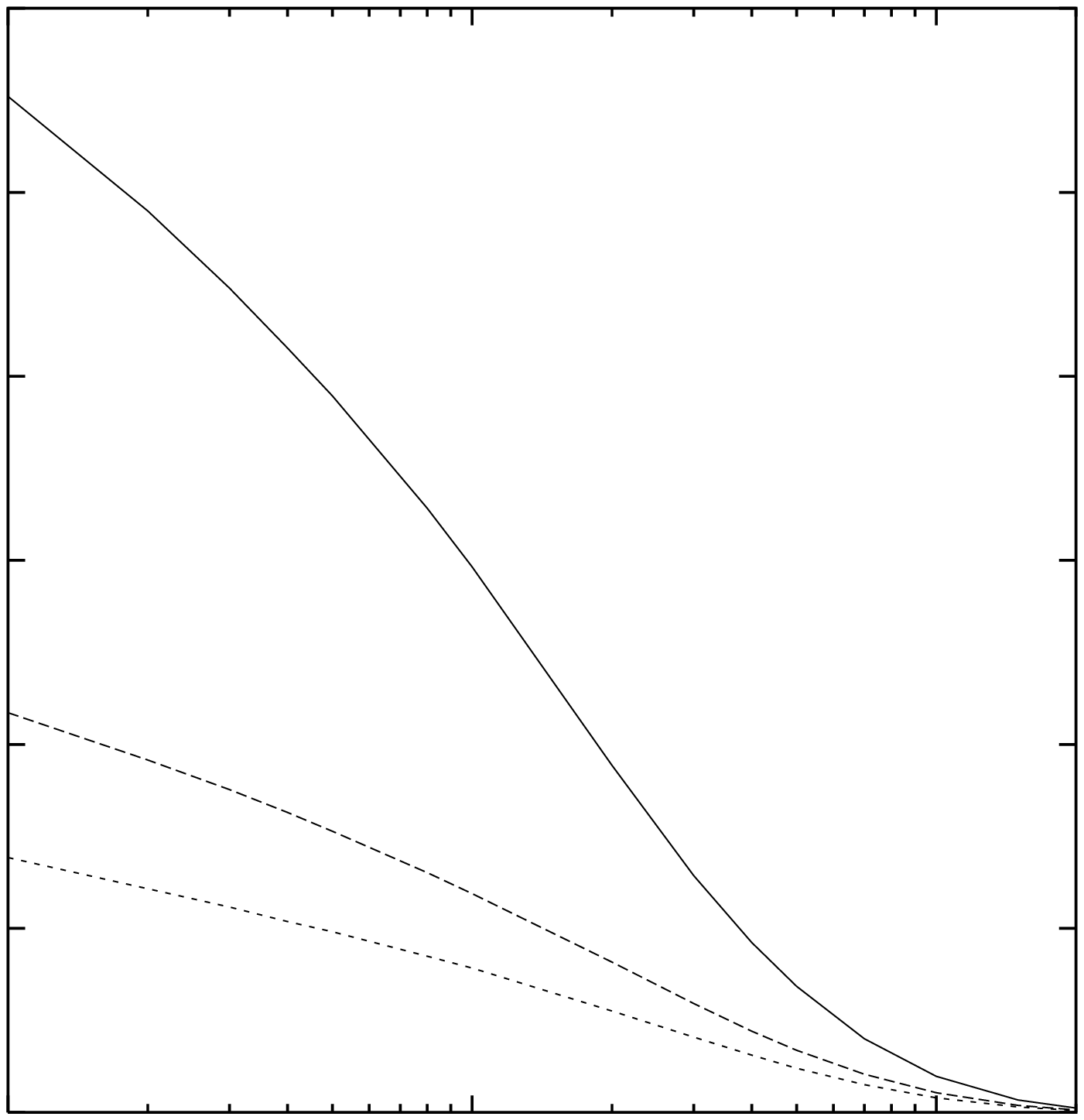}}
\put(85,1){\small $10^{-3}$}
\put(117.5,1){\small $10^{-2}$}
\put(149,1){\small $10^{-1}$}
\put(141,0){$y^J$}
\put(87,82.6){\makebox(0,0)[r]{\small $6$}}
\put(87,69.1){\makebox(0,0)[r]{\small $5$}}
\put(87,56.4){\makebox(0,0)[r]{\small $4$}}
\put(87,43.7){\makebox(0,0)[r]{\small $3$}}
\put(87,31){\makebox(0,0)[r]{\small $2$}}
\put(87,18.5){\makebox(0,0)[r]{\small $1$}}
\put(87,6){\makebox(0,0)[r]{\small $0$}}
\put(93,73){$y^{\gamma} = $}
\put(93,62.5){0.01}
\put(93,32){0.05}
\put(93,17.5){0.1}
\put(110,70){$\sigma(\gamma+(2+1)$-jets) [pb]}
\put(124,-3){b)}
\end{picture}
\caption{\it \gjo\ and \gjt\ cross sections as a function of jet cut and
  photon isolation.}
\label{fig8}
\end{figure}

The rate of \gjt\ events, 
\begin{equation}
R_{\gamma,2+1} = \frac{\sigma(\gamma+(2+1){\rm
    -jets})}{\sigma(\gamma+(1+1){\rm -jets}) + \sigma(\gamma+(2+1){\rm
    -jets})}, 
\end{equation}
increases towards smaller values of $y^J$ (see Fig.\ \ref{fig7}) and
becomes equal to the \gjo\ rate at $y^J \lsim 10^{-3}$, the precise
value depending on $y^{\gamma}$. The dependence on $y^{\gamma}$ is
weaker; in particular, for $y^{\gamma} \gsim 0.02$ the \gjt\ rate is
almost independent on $y^{\gamma}$. The reduction of the cross
sections with increasing $y^{\gamma}$ is stronger for \gjt\ events at
large values of $y^J$ than at small $y^J$, relative to the \gjo\ cross
section, {\it i.e.}\ the ratio $R_{\gamma,2+1}$ increases with
increasing $y^{\gamma}$ at large $y^J$ whereas it decreases with
increasing $y^{\gamma}$ at small $y^J$. For completeness we present
the $y$-cut dependence of the \gjo\ and \gjt\ cross sections in Fig.\ 
\ref{fig8}. Note that also the total cross section has a dependence on
$y^J$, as can be seen from the sum of the results shown in Figs.\ 
\ref{fig8}a and b. The jet algorithm not only defines the classification 
of the hadronic final state into \gjo\ or \gjt\ events, but also affects 
the overall phase space boundaries: smaller values of $y^J$ allow the 
jets to be closer to the remnant jet so that the cut against low-$p^T$ 
partons has a stronger effect.

\begin{figure}[htb] 
\unitlength 1mm
\begin{picture}(100,100)
\put(32,-2){\epsfxsize=9.5cm \epsfysize=10cm
\epsfbox{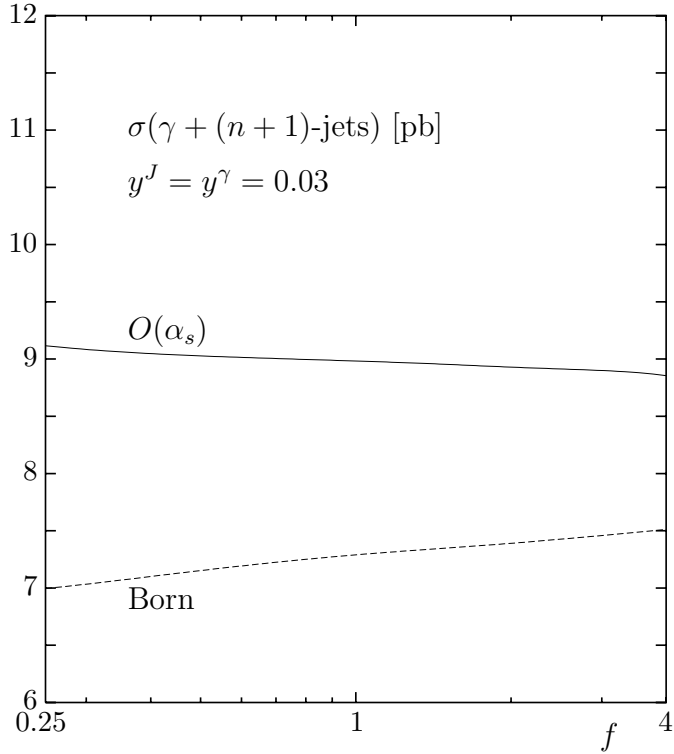}}
\put(52,54){\makebox(0,0)[l]{$O(\alpha_s)$}}
\put(52,18.5){\makebox(0,0)[l]{Born}}
\put(52,81){\makebox(0,0)[l]{$\sigma(\gamma+(n+1)$-jets) [pb]}}
\put(52,74){\makebox(0,0)[l]{$y^J = y^{\gamma} = 0.03$}}
\put(116,0){\makebox(0,0){$f$}}
\put(40.5,2){\makebox(0,0){\small 0.25}}
\put(82.5,2){\makebox(0,0){\small 1}}
\put(123.5,2){\makebox(0,0){\small 4}}
\put(40,96.2){\makebox(0,0)[r]{\small 12}}
\put(40,81){\makebox(0,0)[r]{\small 11}}
\put(40,65.8){\makebox(0,0)[r]{\small 10}}
\put(40,50.6){\makebox(0,0)[r]{\small 9}}
\put(40,35.4){\makebox(0,0)[r]{\small 8}}
\put(40,20.2){\makebox(0,0)[r]{\small 7}}
\put(40,5){\makebox(0,0)[r]{\small 6}}
\end{picture}
\caption{\it Dependence on the renormalization and factorization
  scales $\mu^2_F = \mu^2_R = fQ^2$ of the total cross section ({\it
    i.e.}, the sum of \gjo s and \gjt s) for $y^J = y^{\gamma} = 0.03$.}  
\label{fig9}
\end{figure}

The leading order cross section depends on a factorization scale
$\mu_F$ via the scale entering the parton distribution functions
$q_i(x, \mu_F^2)$. At next-to-leading order, there is an explicit
scale dependence in the \gjo\ cross section through factorization of
the initial state singularities which partly compensates the scale
dependence from the parton distribution functions. In addition, the
explicit factor $\alpha_s$ depends on the renormalization scale. For
simplicity we identify the two scales, which could in principle be
chosen independently from each other.  Figure \ref{fig9} shows the
scale dependence of the leading-order and the next-to-leading order
cross sections where we have used $\mu_F^2 = \mu_R^2 = f Q^2$. Varying
$f$ between $0.25$ and $4$, the total cross section for $ep
\rightarrow e\gamma X$ shows good stability within a few percent. Also
from this figure one can infer that the $O(\alpha_s)$ corrections vary
between 20 and 30\,\%. A 5\,\% measurement of the cross section would
therefore correspond to a 20 to 30\,\% measurement of $\alpha_s$,
assuming negligible uncertainties from parton distribution functions. 

Radiative $ep$ scattering is complementary to usual deep inelastic
scattering since up and down-type quarks contribute with different
weights to the cross sections. In the usual structure function $F_2$,
the sums of $u$- and $d$-type quarks, $U = u + c + \bar{u} + \bar{c}$
and $D = d + s + b + \bar{d} + \bar{s} + \bar{b}$ enter with the
relative factors $e_u^2 : e_d^2 = 4:1$ whereas for the contribution
from quarkonic radiation to $ep \rightarrow e\gamma X$ this ratio is 
$e_u^4 : e_d^4 = 16:1$. In principle, a common analysis of
non-radiative and radiative scattering performed with high enough
precision, would allow a determination of $U$ and $D$ separately. We
therefore investigated the dependence of the total cross section for
$ep \rightarrow e\gamma X$ on the ratio $U/D$. In order to keep the
well-constrained structure function $F_2$ unchanged, we modified the
parton distributions by the following prescription:
\begin{equation}
\begin{array}{l}
\displaystyle
D \rightarrow \delta_d D, \\[1ex]
\displaystyle
U \rightarrow \left( 1 + \frac{1 - \delta_d}{4(U/D)} \right) U. 
\end{array}
\end{equation}
By this the combination $e_u^4 U + e_d^4 D$ is replaced with $\left[1 +
3(1 - \delta_d)/(1 + 16U/D)\right] \times (e_u^4 U + e_d^4 D)$. With 
$U/D \simeq 1.5$, a typical value at $x \simeq 0.1$, one expects a 
12\,\% reduction for $\delta_d = 2$ and a 6\,\% enhancement for 
$\delta_d = 0.5$. In fact, the true change of the cross section is 
smaller ($-5.7$\,\% and $+2.3$\,\% with the cuts (\ref{allcuts})) since
additional contributions from leptonic radiation and quark-lepton
interference, which are not proportional to the fourth power of the
quark charges, are not negligible even in the `signal' region
$\theta_{\gamma} \geq 90^{\circ}$.  It thus seems unlikely that with
respect to a determination of $U/D$ radiative deep inelastic scattering
could become competitive with classical analyses like that of the
difference of proton and neutron cross sections or the $W$ charge
asymmetry in $p\bar{p} \rightarrow W^{\pm} + X$.

%
\section{Summary}

We have described a first next-to-leading order calculation of
isolated photon production in $ep$ scattering at large $Q^2$.  Apart
from providing a sound basis for testing QCD in direct photon
production, our results improve the knowledge of standard model
predictions as a source of background for searches for new physics. We
have discussed numerical results for \gjo\ and \gjt\ cross sections at
HERA. Corresponding measurements will provide valuable information
that will allow to further constrain parton distribution functions, in
particular when combined with results from other experiments. It still
has to be investigated which kinematical variable would be best suited
to obtain the highest sensitivity on the gluon distribution, the $U/D$
ratio, or the strong coupling constant $\alpha_s$. For example, in
photoproduction the distributions with respect to the photon rapidity
or photon transverse momentum (in the HERA laboratory or in the
$\gamma^*p$ center-of-mass frame) turned out to be good choices. One
should also expect that the theoretical uncertainties due to the
parton-level cutoff $y_0^{\gamma}$ could be further reduced by
optimizing the analysis with respect to kinematical cuts, the jet
algorithm ({\it e.g.}, a cone algorithm as used in the study of
$(2+1)$- or $(3+1)$-jet events at HERA \cite{exp-hera}) and modified
photon isolation prescriptions ({\it e.g.}, the so-called
``democratic" clustering procedure \cite{morgan}).

\vspace{6mm}
\begin{flushleft}
{\large \bf Acknowledgment}
\end{flushleft}
We wish to thank D.\ Graudenz for useful discussions.

%

\newpage

\end{document}